\documentclass[tradiabstract]{aa}

\usepackage{url}
\usepackage{natbib}
\usepackage{graphicx}
\usepackage{txfonts}

\bibpunct{(}{)}{;}{a}{}{,} 

\newcommand{\edit}[1]{#1}

\begin{document}

\title{Weeds: a CLASS extension for the analysis of millimeter and
  sub-millimeter spectral surveys}

\author{S. Maret\inst{1} \and P. Hily-Blant\inst{1} \and
  J. Pety\inst{2,3} \and S. Bardeau\inst{2} \and E. Reynier\inst{2}}
\institute{Laboratoire d'Astrophysique de Grenoble, Observatoire de
  Grenoble, Universit\'e Joseph Fourier, CNRS, UMR 571 Grenoble,
  France \and Institut de Radioastronomie Millim\'etrique, 300 Rue de
  la Piscine, 38406 Saint Martin d'H\`eres, France \and LERMA, UMR
  8112, CNRS and Observatoire de Paris, 61 avenue de l'Observatoire,
  75014 Paris, France.}

\date{Received \today; accepted ...}
 
\abstract{The advent of large instantaneous bandwidth receivers and
  high spectral resolution spectrometers on (sub-)millimeter
  telescopes has opened up the possibilities for unbiased spectral
  surveys. Because of the large amount of data they contain, any
  analysis of these surveys requires dedicated software tools. Here we
  present an extension of the widely used CLASS software that we
  developed to that purpose. This extension, named Weeds, allows for
  searches in atomic and molecular lines databases (e.g. JPL or CDMS)
  that may be accessed over the internet using a virtual observatory
  (VO) compliant protocol. The package permits a quick navigation
  across a spectral survey to search for lines of a given
  \edit{species}. Weeds is also capable of modeling a spectrum, as
  often needed for line identification. We expect that Weeds will be
  useful for analyzing and interpreting the spectral surveys that will
  be done with the HIFI instrument on board Herschel, but also
  observations carried-out with ground based millimeter and
  sub-millimeter telescopes and interferometers, such as
  \edit{IRAM-30m and Plateau de Bure, CARMA, SMA, eVLA, and ALMA.}}


\maketitle

\section{Introduction}


A spectral survey consists in a series of spectra covering a
significant spectral domain. At (sub-)millimeter wavelengths, a
spectral survey typically covers several tens of GHz. Spectral surveys
are generally \edit{referred} to as unbiased if they provide a
complete coverage with a uniform sensitivity.  As such, they allow for
a complete census of the species emitting in that band, and sometimes
for discovery of new interstellar species. In addition, because a
given band often contains many transitions of the same species, the
simultaneous analysis of all these lines provides stringent
constraints on the physical conditions in the emitting gas, such as
the density and temperature. Therefore spectral surveys are very
useful for characterizing both the chemical composition and physical
condition in the observed objects.


Ever since the pioneering work of \citet{Johansson84}, who carried-out
an unbiased spectral survey of the Orion~KL star-forming region and
IRC~+10216 carbon-rich star between 72 and 91~GHz with the Onsala
telescope, many spectral surveys have been carried-out at millimeter
and sub-millimeter wavelengths using ground-based telescopes
\citep[see][for a review]{Herbst09}. Because of the limited
sensitivity of the instruments available at that time, early spectral
surveys were targeted at bright star-forming regions, such as Orion~KL
and Sgr~B2 in the millimeter range. Thanks to the increasing
sensitivity of heterodyne receivers and the availability of
sub-millimeter telescopes, these surveys were later extended to higher
frequencies \cite[e.g.][]{Schilke97a,Schilke01,Comito05} and
carried-out towards fainter young stellar objects
\citep[e.g. NGC~1333~IRAS4 or
IRAS16292-2422;][]{Blake94,vanDishoeck95,Blake95}. \edit{A few
  spectral surveys have been carried with millimeter and
  sub-millimeter interferometers, such as OVRO or the SMA
  \citep[e.g.][]{Blake96b,Beuther06}.} The HIFI instrument
\citep{deGraauw10} onboard the Herschel space observatory
\citep{Pilbratt10} now allows for a complete coverage of the almost
unexplored 480-1250 and 1410-1910~GHz frequency bands. Its large
spectral coverage -- up to 4~GHz instantaneous bandwidth -- and
unprecedented sensitivity in this frequency range enable astronomers
to carry-out spectral surveys over almost 1.5~THz down to the line
confusion limit in a few tens of hours. The first spectral surveys
with this instrument have already given spectacular results
\citep{Bergin10,Ceccarelli10}. Among these, we can cite the richness
of the Orion BN-KL spectrum observed at THz frequencies \citep[see
Fig.~2][]{Bergin10} or the discovery of ND in IRAS16293-2422
\citep{Bacmann10}.


Current \edit{developments} in \edit{(sub-)millimeter instruments}
include an increase in the instantaneous bandwidth of the detection
devices. During the past decade, the instantaneous bandwidth of
tunable heterodyne receivers has increased by more than an order of
magnitude, now routinely reaching $\sim$10 GHz. Other technologies
(e.g. HEMT, FCRAO and IRAM) have already provided several tens of GHz,
although it is still unclear whether the sensitivity of these
receivers can match that of SIS receivers. This increase in bandwidth
has been accomplished in parallel with the advent of digital
spectrometers (autocorrelators, fast Fourier transform), the
versatility of which allow the coverage of such bandwidth with a
spectral resolution down to a few hundred kHz. As a result, unbiased
spectral surveys of the 3 mm atmospheric window ($\nu = 80 - 117$~GHz)
can be done with the IRAM-30m telescope in $\sim$10~hours, with a 2~mK
noise at 1$\sigma$ in 2~MHz ($\sim$6~km/s) spectral channels.  The
ALMA interferometer will also permits coverage of large frequency
windows, providing spectral cubes with up to 8~GHz bandwidth
\citep{Wooten08}. \edit{Thanks to its sensitivity, this instrument
  will allow, in its compact configuration, line surveys to be
  carried-out down to the confusion limit toward a large number of
  sources}. Spectral surveys are thus still in their infancy and will
very likely become routine observing modes in the coming years.


Spectral surveys covering large frequency bands require specific tools
to be analyzed efficiently. In this article, we present a software
that is intended for the analysis of spectral surveys. In
\S~\ref{sec:spectr-surv-analys}, we briefly describe how such surveys
are analyzed. In \S~\ref{sec:weeds-design-impl} we detail how our
software was designed and implemented to carried-out such an
analysis. Finally \S~\ref{sec:concl-prosp} concludes this article and
discuss future developments.

\section{Spectral surveys analysis}
\label{sec:spectr-surv-analys}


The analysis of a spectral survey usually consists in identifying the
various lines and in deriving the physical and chemical properties of
the emitting gas (density, temperature and column densities of the
observed species). The main difficulty in such identification is that
large molecules may have hundreds of lines in the (sub-)millimeter
range. \edit{These species -- such as methanol, methyl formate or
  dimethyl ether -- are often named \emph{weeds} by spectroscopists.}
If the lines are too broad, they may overlap and blend together, which
makes the identification of weaker lines difficult. This is the
\emph{line confusion limit} \citep{Schilke97a}: line identification is
not limited by the signal-to-noise of the observations, but by the
line blending.

Because of this problem, extreme care must be taken when identifying
species from a spectral survey. \citet{Herbst09} summarize the
criteria for a firm detection as follows: ``\emph{(i)} Rest
frequencies are accurately known to 1:10$^{7}$, either from direct
laboratory measurements or from a high-precision Hamiltonian model;
\emph{(ii)} observed frequencies of clean, nonblended lines agree with
rest frequencies for a single well-determined velocity of the source;
if a source has a systematic velocity field as determined from simple
molecules, any velocity gradient found for lines of a new complex
molecule cannot be a random function of transition frequency;
\emph{(iii)} all predicted lines of a molecule based on an LTE
spectrum at a well-defined rotational temperature and appropriately
corrected for beam dilution are present in the observed spectrum at
roughly their predicted relative intensities. A single anticoincidence
(that is, a predicted line missing in the observational data) is a
much stronger criterion for rejection than hundreds of coincidences
are for identification. This last criterion is one of the strongest
arguments for complete line surveys rather than targeted line
searches.''

The rest frequencies needed to fulfill criterion \emph{(i)} are
usually taken from spectral lines catalogs, such as the Cologne
Database for Molecular Spectroscopy \citep[CDMS;][]{Muller01} or the
JPL Molecular Spectroscopy catalog \citep{Pickett98}.  For criterion
\emph{(ii)}, we need to compare the consistency of the centroid
velocities of all the line candidates.  Finally criterion \emph{(iii)}
requires to perform a model of the predicted emission of the given
species so that it can be compared with the observations. The
traditional technique for this consist in building a rotational
diagram \citep{Goldsmith99} to see if all detected lines agree with a
single rotational temperature and column density. Alternatively, one
can compute synthetic spectrum and compare it directly with the
observations -- a technique called \emph{forward fitting}
\citep{Comito05}. This approach is also extremely useful when one
wants to search for weak lines of a specie among hundreds from various
weeds: a synthetic spectrum of the emission of the weeds can be
constructed to fit the observed transitions in an iterative
fashion. \edit{Once the brightest lines have been modeled, one can
  compare the synthetic spectrum to the observed one to look for lines
  from less abundant species \citep[see ][for an example of this
  technique]{Belloche08}.}  Of course, this also allows the physical
and chemical properties of the emitting gas to be derived.


Since spectral surveys may contain thousands of lines, they require
specific tools to be efficiently analyzed. Two packages have been
developed for that purpose. The first of them, XCLASS
\citep{Schilke01}, is an extension of the widely used CLASS data
reduction software, which is part of Gildas. XCLASS contains a
spectral line database which is built from the CDMS and JPL
catalogs. Technically, it uses the MySQL database server which must be
installed on the user computer. This database may be updated manually,
by replacing the database file by the one provided by the program
authors. XCLASS allows the user to look for lines corresponding to a
given frequency in its catalog, but also to make a model at the LTE of
the observed spectra.
XCLASS has been successfully used to reduce several spectral surveys
obtained with the CSO and the IRAM-30m
\citep{Schilke01,Comito05,Belloche08}. However, XCLASS is based on an
obsolete version of CLASS, which is not maintained
\edit{anymore}. Indeed, the CLASS internal structures was largely
rewritten in 2005-2006 to adapt to the challenges of data reductions
coming with the recent generation of receivers
\citep{HilyBlant05}. The second package, CASSIS, has been developed
primarily to analyze Herschel-HIFI spectral surveys, although it can
be used to analyze surveys from ground based telescopes as
well. CASSIS itself does not have data reduction capabilities;
therefore data must first be reduced in another software such as CLASS
or HIPE (Ott et al., in prep.) before analysis in CASSIS. CASSIS uses
a database which is built from the CDMS and the JPL catalog; in recent
CASSIS versions, this database (SQLite) is embedded in the program so
that an external database server is no longer required.  Like XCLASS,
CASSIS allows the forward-fitting of a spectrum, but also the search
for the various transitions of a given specie.

\section{Weeds design and implementation}
\label{sec:weeds-design-impl}

\subsection{General design}
\label{sec:general-design}


Weeds has been designed specifically to analyze spectral surveys,
following the approach presented in
\S~\ref{sec:spectr-surv-analys}. Although its development was inspired
by the XCLASS and CASSIS packages, it is different in several
aspects. Weeds is an extension of the current version of the CLASS
software, and is mostly written in Python language, except for a few
command written in the Gildas command interpreter (SIC) language. To
do this, Weeds uses the new possibility offered by GILDAS to
interleave Python and SIC in the same session \citep{Bardeau10}. In
particular, the variable contents are shared between Python and
SIC. Python has several advantages over other languages for developing
such extensions. It benefits from a large library of modules that
allow complex tasks -- such as making a query in a VO-compliant
database, see \S~\ref{sec:spectr-line-catal} -- to be done relatively
easily. Although it is interpreted, it is still computationally
efficient, because critical modules (e.g. the module for array
computations that we use for the spectra modeling, see
\S~\ref{sec:spectra-modeling}) are written in compiled languages such
as C or Fortran.  Weeds is distributed with Gildas since April
2010. The source code is freely available from the IRAM
website\footnote{\url{http://iram.fr/IRAMFR/GILDAS/}}. A user manual
is also available on that page.


Because Weeds is an extension of CLASS, it can be used to analyze any
data format that CLASS supports. In practice, the CLASS data format is
used by many ground-based telescopes (e.g. IRAM-30m, CSO and
APEX). Data from other telescopes can be converted to FITS format and
imported into CLASS as well. For example, Herschel-HIFI can be
imported into CLASS through the FITS filler delivered by the HIPE data
reduction software (Delforge et al., in prep.). In order to analyze
data in Weeds, the data must have been calibrated and reduced
first. The reduction usually consists in flagging the bad channels,
averaging the scan covering the same frequency range together, and
removing a polynomial baseline. If the data were obtained with double
sideband (DSB) receiver, \edit{sideband deconvolution might be needed
  in order to produce a SSB spectrum. This requires a special
  observing technique, i.e. a number of overlapping spectra with
  shifted local oscillator frequency.  Deconvolution can then be
  performed in CLASS using the algorithm developed by
  \citet{Comito02}}. Thus data reduction and analysis can be done
within the same environment.


\subsection{Spectral line catalogs queries}
\label{sec:spectr-line-catal}


As mentioned above, line identification requires repeated queries to
spectral lines catalogs, such as the CDMS or the JPL. Unlike XCLASS
and CASSIS -- who require a custom catalog installed on the user's
computer -- Weeds performs queries in spectral line databases through
the Internet\footnote{However, Weeds can make a cache of part or an
  entire catalog, so that it can be used later with no Internet
  connection.}. This has the advantage of not requiring any update of
a custom catalog: changes in the database, such as species addition or
line frequency corrections or updates, are readily available in
Weeds. In order to make queries in spectral lines catalogs, we have
implemented the VO-compliant \emph{Simple Line Access Protocol}
\citep[SLAP;][]{Salgado09} in Weeds. This protocol allows spectral
line databases queries to be made in a standardized way; any database
that implements the protocol can be accessed by Weeds. Because it is a
VO standard, it is likely that more and more spectral line database
will use it in the future. Nonetheless, as of this writing only the
CDMS is accessible using that protocol, through an interface at the
Paris VO Observatory \citep{Moreau08}. Therefore, in order to access
the JPL catalog from Weeds, we have implemented queries in the
specific protocol which is used by this database. The CDMS can be
accessed through its own protocol as well.

\edit{For the moment, only one database can be used at a time; it is
  not possible to combine the catalogs, i.e. to use species some out
  the JPL and some out the CDMS. In the future, the VAMDC
  project\footnote{\url{http://www.vamdc.org/}} will provide a
  single, unified database, including state-of-art spectroscopic data
  from both the CDMS and the JPL catalogs. We plan on implementing an
  access to this database from Weeds as soon as it it released.}

From the user point of view, Weeds provides a command to search for
lines corresponding to a given frequency range in a spectral line
catalog. The user can select a region on the spectrum displayed in
CLASS, and the command prints all the lines from the catalog around
the region selected. The lines can be filtered out on the basis of the
species they belong to, their Einstein coefficient, or their upper
level energy. For double sideband spectra, a command option allows the
search for lines from the image band.

\subsection{Lines browsing/identification}
\label{sec:lines-brows}


To secure the detection of a species in a spectral survey, one needs,
according to criterion \emph{(ii)} to search for all the transitions
of that species in the entire frequency range covered by the
survey. One also needs to measure the velocity of each line to check
that they correspond to a single velocity. Weeds allows the user to
browse through a survey rapidly. For this, Weeds has a command to
search for all the lines of a given specie that fall in the frequency
range covered by the survey. The command prints the lines in the
terminal, but also builds an internal index containing all these
lines, that we can order either by increasing frequency or increasing
upper level energy. Another command allow the user to examine each of
the line candidate one by one, to see if the line is detected or
not. This command makes a zoom on a small frequency region around the
(expected) line, and also sets the velocity scale with respect to the
rest frequency of the line. A vertical mark is also drawn on the
displayed spectrum at the source velocity, so that we can easily
determine if the line is detected or not. A Gaussian fit of the
observed line may be performed to determine the velocity of each line.

\subsection{Spectra modeling}
\label{sec:spectra-modeling}


Once several transitions of a given specie have been found, one needs
to check if the relative intensities of these components agree with a
single excitation temperature (criterion \emph{(iii)}). In addition,
one needs to make sure that non-detected lines are consistent with the
excitation temperature derived from other species -- or in other
words, that no lines are ``missing''. For this Weeds allows the user
to compute a synthetic spectrum that can be compared directly with the
observations (forward-fitting). Following the approach used in XCLASS
and described in \citet{Comito05} the synthetic spectrum is computed
assuming \edit{that the emission arises from one or several components
  at the LTE. Although this approximation is simplistic -- it is well
  known that in the interstellar medium species are often out of local
  thermodynamic equilibrium, and many sources are known to have
  density and temperature gradients -- yet such a zeroth-order
  approach is often extremely useful to identify lines, as mentioned
  above. Once the lines have been identified, a more realistic
  modeling, taking into account non-LTE excitation effects as well as
  the source structure, can be carried-out.}

\edit{Under these assumptions,} and after baseline subtraction, the
brightness temperature of a given species as a function of the rest
frequency $\nu$ is given by:


\begin{equation}
  \label{eq:1}
  T_\mathrm{B} \left( \nu \right) = \eta \left[ J_\nu \left( T_\mathrm{ex}
    \right) - J_\nu \left( T_\mathrm{bg} \right) \right] \left( 1 - e^{-\tau
      \left( \nu \right)} \right)
\end{equation}

\noindent
where $\eta$ is beam dilution factor, which, for a source with a
Gaussian brightness profile and a Gaussian beam, is equal to:

\begin{equation}
  \eta = \frac{\theta^{2}_{s}}{\theta^{2}_{s} + \theta^{2}_{t}}
\end{equation}

\noindent
where $\theta_{s}$ and $\theta_{t}$ the source and telescope beam FWHM
sizes, respectively. For a sake of simplicity, the latter is assumed
to be given by the diffraction limit\footnote{(Sub-)millimeter
  telescopes usually have tapers that limit the power received in
  side-lobes. Because of this, the telescope beam size may be
  different that of a purely diffraction limited antenna of the same
  diameter. However, the difference between the two is usually small:
  at 100~GHz, the measured FWHM of the IRAM-30m is 24.6\arcsec, while
  Eq.\ref{eq:2} gives 25.2\arcsec.}

\begin{equation}
  \label{eq:2}
  \theta_{t} = 1.22 \frac{c}{\nu D}
\end{equation}

\noindent
where $c$ is the light speed and $D$ is the diameter of the
telescope. $T_\mathrm{bg}$ is \edit{the brightness temperature of the
  background emission, i.e. the physical temperature that would have a
  black body producing the same background continuum emission
  (e.g. 2.73~K for the cosmic microwave background).}  $T_\mathrm{ex}$
is the excitation temperature, and the opacity $\tau \left( \nu
\right)$ is:

\begin{equation}
  \label{eq:3}
  \tau \left( \nu \right) = \frac{c^{2}}{8 \pi \nu^2} 
  \frac{N_\mathrm{tot}}{Q(T_\mathrm{ex})} \sum_i A^{i} \,
  g^{i}_{u} \, e^{-E^{i}_{u} / k T_\mathrm{ex}}
  \left( e^{h \nu^{i}_{0} / k T_\mathrm{ex}} - 1 \right) \phi^{i}
\end{equation}

\noindent
where the summation is over each line of the considered species. Here
$N_\mathrm{tot}$ is the total column density of the species
considered, $Q(T_\mathrm{ex})$ is the partition function, $A^{i}$ is
the Einstein coefficient of the $i$ line, $g^{i}_{u}$ and $E_{u}$ are
the upper level degeneracy and energy of the $i$ line, and $\phi^{i}$
is the $i$ line profile function. The latter is given by:

\begin{equation}
  \phi^{i} = \frac{1}{\sigma \sqrt{2 \pi}}
    e^{- \left( \nu - \nu^{i}_{0} \right)^{2} / 2 \sigma^{2}}
\end{equation}

\noindent
where $\nu^{i}_0$ the is $i$ line rest frequency and $\sigma$ the line
width in frequency units at $1/e$. $\sigma$ can be expressed as a
function of the line FWHM in velocity units $\Delta{V}$ as follows:

\begin{equation}
  \sigma = \frac{\nu^{i}_0}{c \sqrt{8 \ln 2}} \Delta{V}
\end{equation}

\noindent
\edit{Note that some of the model parameters may be degenerate in
  certain cases. In the optically thick or thin limits, the source
  size and temperature or the size and column density are degenerate,
  respectively (see Eq. \ref{eq:1} and \ref{eq:2}). This degeneracy
  can be usually lifted if both thick and thin lines are present in
  the survey, or if lines from an rare isotopologue are detected
  together with the main one (e.g. $^{13}$CH$_{3}$OH and
  CH$_{3}$OH). The source size may also be constrained from
  interferometric observations.}
 
Several components with e.g. different kinetic temperature or column
density can be included in the computation. For this, we assume that
the various components are not coupled radiatively -- that is a photon
from one component can not be absorbed by a another, foreground
component -- in which case the emerging spectrum is simply the sum of
the brightness temperature of each components given by
Eq.~(\ref{eq:1}). Each of these component can be Doppler-shifted with
respect to each other, which is useful when modeling sources with
several components at different velocities. \edit{It is also possible
  to compute the spectra from several species; this is done} by a
summation of Eq.~(\ref{eq:1}) over each specie.

The column densities, kinetic temperatures, Doppler width and source
sizes for each species and components are read from a text
file. Einstein coefficients, upper level degeneracy and energies as
well as the partition functions are taken from spectral line
catalogs. Because these catalogs usually give the partition functions
at a few temperatures only, the partition function at the user
temperature is computed from a linear interpolation (or extrapolation
if the user given temperature is outside the range of temperature
provided in the catalog). When computing the synthetic spectrum, a
frequency sampling corresponding to the minimum $\Delta{V}$ divided by
10 is taken (or a frequency sampling equal to that of the observed
spectra, if it smaller than the minimum $\Delta{V}$ divided by
10). This ensures that the sampling at all frequencies and for all
species and components is sufficient. At the end of the computation,
the synthetic spectrum is re-sampled to the same channel spacing than
the observed spectrum in order to take the channel dilution factor
into account. This allows for a direct comparison between the
synthetic and observed spectra.

\begin{figure*}
  \centering
  \includegraphics[angle=-90,width=\textwidth]{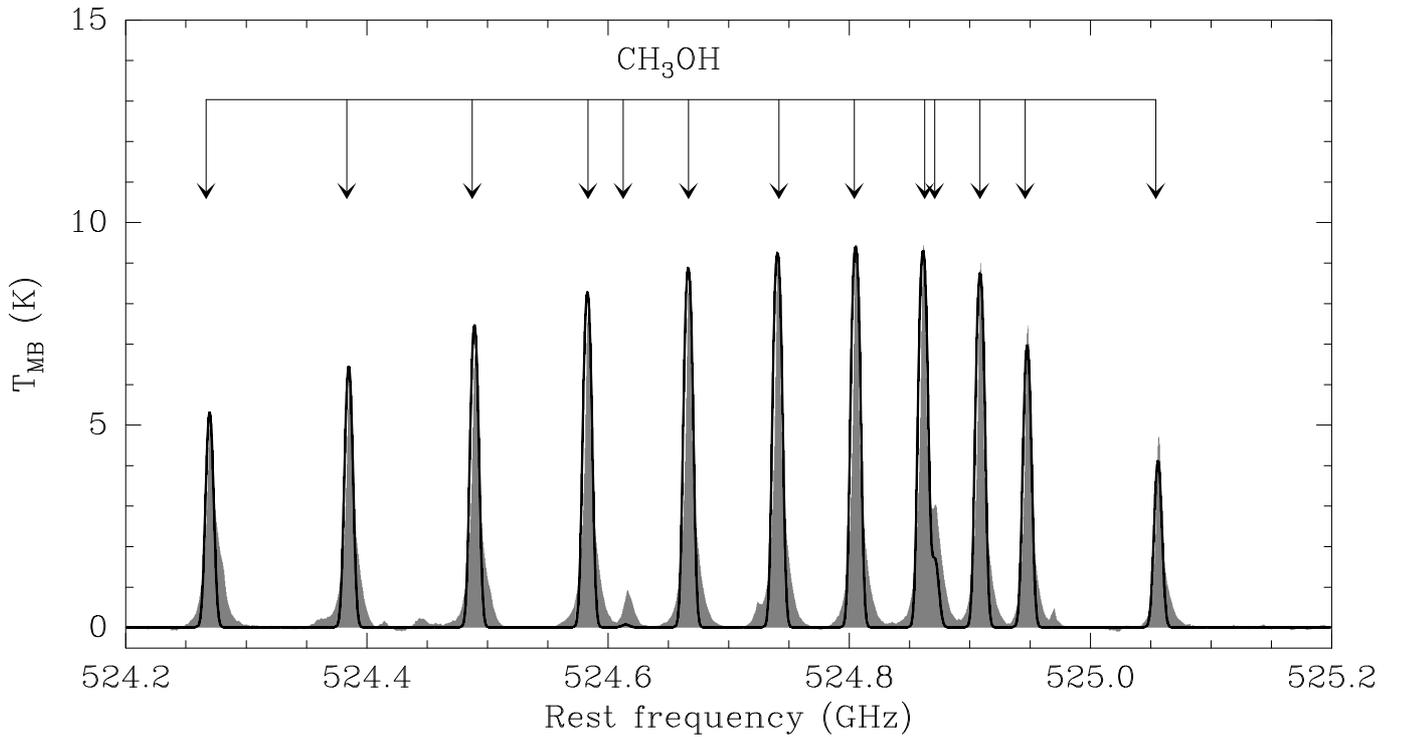}
  \caption{Spectra between 524.2 and 525.5~GHz observed towards
    Orion-KL with \emph{Herschel-HIFI} (filled histogram) and LTE
    model produced with Weeds (continuous black line). The rest
    frequencies of several detected methanol lines are indicated.}
  \label{fig:ch3oh-orionkl}
\end{figure*}

On Fig.~\ref{fig:ch3oh-orionkl}, we show an example of such a
modeling. The figure shows a spectrum between 524.2 and 525.2~GHz
observed towards Orion-KL with \emph{Herschel-HIFI} as part of the
HEXOS guaranteed time key program \citep{Bergin10}. These data have
been already presented by \citet{Wang10}. Several methanol lines are
detected. On this figure we show model predictions computed with a
Weeds for a single component source with $N(\mathrm{CH_{3}OH}) = 2
\times 10^{17}$~cm$^{-2}$, $\theta_S = 18\arcsec$, $T = 80$~K and
$\Delta V = 4$~km/s, and using the JPL database. \edit{Overall, the
  model predictions are in good agreement with the observations -- in
  particular, we reproduce successfully the relative intensity of the
  brightest lines. On the other hand, this simple model underestimates
  the small line at 524620 MHz and the shoulder at 524880 MHz, maybe
  suggesting several emitting components and/or non-LTE
  excitation. Note that for the given parameters the emission is
  predicted to be optically thin, so that the column density and
  source size can not be constrained independently. A complete
  analysis of the methanol emission in this source is clearly beyond
  the scope of this paper; however this example demonstrates how a
  simple LTE model is useful to identify lines in a spectral survey.}
Finally, we have crossed-checked these model predictions with CASSIS
and two packages were found to be in excellent agreement.

\section{Conclusions and prospects}
\label{sec:concl-prosp}

We have presented an extension of the CLASS data reduction software
for analyzing spectral surveys. This extension allows the user to make
queries in spectral line databases using a VO compliant protocol. It
also allows the user to quickly search for the various transitions of
a given specie. Finally it can compute model predictions at the LTE,
as often needed to identify lines in spectra close to the confusion
limit. Weeds has already been successfully used to analyze part of the
IRAS~16293-2422 survey obtained with \emph{Herschel-HIFI}
\citep{Bacmann10,HilyBlant10}, and we expect that it will be useful
for future spectral surveys with this instrument as well. We think
that it will become a standard tool for analyzing spectral surveys
obtained with single dish ground based telescopes such as the
IRAM-30m. Yet, Weeds is not limited to the analysis of single dish
observations. \edit{It may be used to analyze spectral surveys
  obtained with interferometers as well, such as the IRAM Plateau de
  Bure, CARMA, the SMA, and the upcoming ALMA and eVLA
  interferometers.  In fact, since} Weeds is written in Python, it
could be used from the Python based CASA software, that will be used
by the eVLA and ALMA.  However, analyzing ALMA data will be
challenging, because these data will consist in large spectral cubes,
i.e. essentially a spectral survey on large number of pixels. In fact,
doing such an analysis by hand, i.e. identifying the various
lines/species on each spectrum of map is probably impossible; this
will require some automatic fitting tools to extract the relevant
information (column densities and excitation temperature of the
various species) as a function of position. Such tools require
efficient minimization algorithms to fit a model with a large number
of free parameters to the data. The development of such tools is
already in progress \edit{(e.g. in XCLASS using the MAGIX minimization
  framework), and implementing these automatic fitting tools in Weeds
  would be desirable in the future.}

\begin{acknowledgements}
  The authors would like to thank Peter Schilke and Emmanuel Caux for
  fruitful discussions on the analysis of spectral surveys. We are
  also grateful to Charlotte Vastel for helping us testing the LTE
  modeling done in Weeds against CASSIS, and to Shyia Wang for
  providing us the Orion-KL spectrum prior to publication. Finally, we
  wish to thank the persons in charge of maintaining and the CDMS, JPL
  and Paris VO databases; without their continuous efforts, the
  development of analysis software such as Weeds would not be
  possible. We are especially grateful to Holger M\"u ller, Brian
  Drouin and Nicolas Moreau for their help in implementing access to
  these databases in Weeds.
\end{acknowledgements}

\bibliographystyle{aa}
\bibliography{bibliography}

\begin{thebibliography}{27}
\expandafter\ifx\csname natexlab\endcsname\relax\def\natexlab#1{#1}\fi

\bibitem[{{Bacmann} {et~al.}(2010){Bacmann}, {Caux}, {Hily-Blant}, {Parise},
  {Pagani}, {Bottinelli}, {Maret}, {Vastel}, {Ceccarelli}, {Cernicharo},
  {Henning}, {Castets}, {Coutens}, {Bergin}, {Blake}, {Crimier}, {Demyk},
  {Dominik}, {Gerin}, {Hennebelle}, {Kahane}, {Klotz}, {Melnick}, {Schilke},
  {Wakelam}, {Walters}, {Baudry}, {Bell}, {Benedettini}, {Boogert}, {Cabrit},
  {Caselli}, {Codella}, {Comito}, {Encrenaz}, {Falgarone}, {Fuente},
  {Goldsmith}, {Helmich}, {Herbst}, {Jacq}, {Kama}, {Langer}, {Lefloch}, {Lis},
  {Lord}, {Lorenzani}, {Neufeld}, {Nisini}, {Pacheco}, {Pearson}, {Phillips},
  {Salez}, {Saraceno}, {Schuster}, {Tielens}, {van der Tak}, {van der Wiel},
  {Viti}, {Wyrowski}, {Yorke}, {Faure}, {Benz}, {Coeur-Joly}, {Cros},
  {G{\"u}sten}, \& {Ravera}}]{Bacmann10}
{Bacmann}, A., {Caux}, E., {Hily-Blant}, P., {et~al.} 2010, \aap, 521, L42+

\bibitem[{Bardeau {et~al.}(2010)Bardeau, Reynier, Pety, \&
  Guilloteau}]{Bardeau10}
Bardeau, S., Reynier, E., Pety, J., \& Guilloteau, S. 2010, PYGILDAS:
  Interleaving Python and GILDAS, Tech. rep., IRAM

\bibitem[{{Belloche} {et~al.}(2008){Belloche}, {Menten}, {Comito},
  {M{\"u}ller}, {Schilke}, {Ott}, {Thorwirth}, \& {Hieret}}]{Belloche08}
{Belloche}, A., {Menten}, K.~M., {Comito}, C., {et~al.} 2008, \aap, 482, 179

\bibitem[{{Bergin} {et~al.}(2010){Bergin}, {Phillips}, {Comito}, {Crockett},
  {Lis}, {Schilke}, {Wang}, {Bell}, {Blake}, {Bumble}, {Caux}, {Cabrit},
  {Ceccarelli}, {Cernicharo}, {Daniel}, {de Graauw}, {Dubernet},
  {Emprechtinger}, {Encrenaz}, {Falgarone}, {Gerin}, {Giesen}, {Goicoechea},
  {Goldsmith}, {Gupta}, {Hartogh}, {Helmich}, {Herbst}, {Joblin}, {Johnstone},
  {Kawamura}, {Langer}, {Latter}, {Lord}, {Maret}, {Martin}, {Melnick},
  {Menten}, {Morris}, {M{\"u}ller}, {Murphy}, {Neufeld}, {Ossenkopf}, {Pagani},
  {Pearson}, {P{\'e}rault}, {Plume}, {Roelfsema}, {Qin}, {Salez}, {Schlemmer},
  {Stutzki}, {Tielens}, {Trappe}, {van der Tak}, {Vastel}, {Yorke}, {Yu}, \&
  {Zmuidzinas}}]{Bergin10}
{Bergin}, E.~A., {Phillips}, T.~G., {Comito}, C., {et~al.} 2010, \aap, 521,
  L20+

\bibitem[{{Beuther} {et~al.}(2006){Beuther}, {Zhang}, {Reid}, {Hunter},
  {Gurwell}, {Wilner}, {Zhao}, {Shinnaga}, {Keto}, {Ho}, {Moran}, \&
  {Liu}}]{Beuther06}
{Beuther}, H., {Zhang}, Q., {Reid}, M.~J., {et~al.} 2006, \apj, 636, 323

\bibitem[{{Blake} {et~al.}(1996){Blake}, {Mundy}, {Carlstrom}, {Padin},
  {Scott}, {Scoville}, \& {Woody}}]{Blake96b}
{Blake}, G.~A., {Mundy}, L.~G., {Carlstrom}, J.~E., {et~al.} 1996, \apjl, 472,
  L49+

\bibitem[{{Blake} {et~al.}(1995){Blake}, {Sandell}, {van Dishoeck},
  {Groesbeck}, {Mundy}, \& {Aspin}}]{Blake95}
{Blake}, G.~A., {Sandell}, G., {van Dishoeck}, E.~F., {et~al.} 1995, \apj, 441,
  689

\bibitem[{{Blake} {et~al.}(1994){Blake}, {van Dishoek}, {Jansen}, {Groesbeck},
  \& {Mundy}}]{Blake94}
{Blake}, G.~A., {van Dishoek}, E.~F., {Jansen}, D.~J., {Groesbeck}, T.~D., \&
  {Mundy}, L.~G. 1994, \apj, 428, 680

\bibitem[{{Ceccarelli} {et~al.}(2010){Ceccarelli}, {Bacmann}, {Boogert},
  {Caux}, {Dominik}, {Lefloch}, {Lis}, {Schilke}, {van der Tak}, {Caselli},
  {Cernicharo}, {Codella}, {Comito}, {Fuente}, {Baudry}, {Bell}, {Benedettini},
  {Bergin}, {Blake}, {Bottinelli}, {Cabrit}, {Castets}, {Coutens}, {Crimier},
  {Demyk}, {Encrenaz}, {Falgarone}, {Gerin}, {Goldsmith}, {Helmich},
  {Hennebelle}, {Henning}, {Herbst}, {Hily-Blant}, {Jacq}, {Kahane}, {Kama},
  {Klotz}, {Langer}, {Lord}, {Lorenzani}, {Maret}, {Melnick}, {Neufeld},
  {Nisini}, {Pacheco}, {Pagani}, {Parise}, {Pearson}, {Phillips}, {Salez},
  {Saraceno}, {Schuster}, {Tielens}, {van der Wiel}, {Vastel}, {Viti},
  {Wakelam}, {Walters}, {Wyrowski}, {Yorke}, {Liseau}, {Olberg}, {Szczerba},
  {Benz}, \& {Melchior}}]{Ceccarelli10}
{Ceccarelli}, C., {Bacmann}, A., {Boogert}, A., {et~al.} 2010, \aap, 521, L22+

\bibitem[{{Comito} \& {Schilke}(2002)}]{Comito02}
{Comito}, C. \& {Schilke}, P. 2002, \aap, 395, 357

\bibitem[{{Comito} {et~al.}(2005){Comito}, {Schilke}, {Phillips}, {Lis},
  {Motte}, \& {Mehringer}}]{Comito05}
{Comito}, C., {Schilke}, P., {Phillips}, T.~G., {et~al.} 2005, \apjs, 156, 127

\bibitem[{{de Graauw} {et~al.}(2010){de Graauw}, {Helmich}, {Phillips},
  {Stutzki}, {Caux}, {Whyborn}, {Dieleman}, {Roelfsema}, {Aarts}, {Assendorp},
  {Bachiller}, {Baechtold}, {Barcia}, {Beintema}, {Belitsky}, {Benz}, {Bieber},
  {Boogert}, {Borys}, {Bumble}, {Ca{\"i}s}, {Caris}, {Cerulli-Irelli},
  {Chattopadhyay}, {Cherednichenko}, {Ciechanowicz}, {Coeur-Joly}, {Comito},
  {Cros}, {de Jonge}, {de Lange}, {Delforges}, {Delorme}, {den Boggende},
  {Desbat}, {Diez-Gonz{\'a}lez}, {di Giorgio}, {Dubbeldam}, {Edwards},
  {Eggens}, {Erickson}, {Evers}, {Fich}, {Finn}, {Franke}, {Gaier}, {Gal},
  {Gao}, {Gallego}, {Gauffre}, {Gill}, {Glenz}, {Golstein}, {Goulooze},
  {Gunsing}, {G{\"u}sten}, {Hartogh}, {Hatch}, {Higgins}, {Honingh}, {Huisman},
  {Jackson}, {Jacobs}, {Jacobs}, {Jarchow}, {Javadi}, {Jellema}, {Justen},
  {Karpov}, {Kasemann}, {Kawamura}, {Keizer}, {Kester}, {Klapwijk}, {Klein},
  {Kollberg}, {Kooi}, {Kooiman}, {Kopf}, {Krause}, {Krieg}, {Kramer},
  {Kruizenga}, {Kuhn}, {Laauwen}, {Lai}, {Larsson}, {Leduc}, {Leinz}, {Lin},
  {Liseau}, {Liu}, {Loose}, {L{\'o}pez-Fernandez}, {Lord}, {Luinge}, {Marston},
  {Mart{\'{\i}}n-Pintado}, {Maestrini}, {Maiwald}, {McCoey}, {Mehdi}, {Megej},
  {Melchior}, {Meinsma}, {Merkel}, {Michalska}, {Monstein}, {Moratschke},
  {Morris}, {Muller}, {Murphy}, {Naber}, {Natale}, {Nowosielski}, {Nuzzolo},
  {Olberg}, {Olbrich}, {Orfei}, {Orleanski}, {Ossenkopf}, {Peacock}, {Pearson},
  {Peron}, {Phillip-May}, {Piazzo}, {Planesas}, {Rataj}, {Ravera}, {Risacher},
  {Salez}, {Samoska}, {Saraceno}, {Schieder}, {Schlecht}, {Schl{\"o}der},
  {Schm{\"u}lling}, {Schultz}, {Schuster}, {Siebertz}, {Smit}, {Szczerba},
  {Shipman}, {Steinmetz}, {Stern}, {Stokroos}, {Teipen}, {Teyssier}, {Tils},
  {Trappe}, {van Baaren}, {van Leeuwen}, {van de Stadt}, {Visser}, {Wildeman},
  {Wafelbakker}, {Ward}, {Wesselius}, {Wild}, {Wulff}, {Wunsch}, {Tielens},
  {Zaal}, {Zirath}, {Zmuidzinas}, \& {Zwart}}]{deGraauw10}
{de Graauw}, T., {Helmich}, F.~P., {Phillips}, T.~G., {et~al.} 2010, \aap, 518,
  L6+

\bibitem[{{Goldsmith} \& {Langer}(1999)}]{Goldsmith99}
{Goldsmith}, P.~F. \& {Langer}, W.~D. 1999, \apj, 517, 209

\bibitem[{{Herbst} \& {van Dishoeck}(2009)}]{Herbst09}
{Herbst}, E. \& {van Dishoeck}, E.~F. 2009, \araa, 47, 427

\bibitem[{{Hily-Blant} {et~al.}(2010){Hily-Blant}, {Maret}, {Bacmann},
  {Bottinelli}, {Parise}, {Caux}, {Faure}, {Bergin}, {Blake}, {Castets},
  {Ceccarelli}, {Cernicharo}, {Coutens}, {Crimier}, {Demyk}, {Dominik},
  {Gerin}, {Hennebelle}, {Henning}, {Kahane}, {Klotz}, {Melnick}, {Pagani},
  {Schilke}, {Vastel}, {Wakelam}, {Walters}, {Baudry}, {Bell}, {Benedettini},
  {Boogert}, {Cabrit}, {Caselli}, {Codella}, {Comito}, {Encrenaz}, {Falgarone},
  {Fuente}, {Goldsmith}, {Helmich}, {Herbst}, {Jacq}, {Kama}, {Langer},
  {Lefloch}, {Lis}, {Lord}, {Lorenzani}, {Neufeld}, {Nisini}, {Pacheco},
  {Phillips}, {Salez}, {Saraceno}, {Schuster}, {Tielens}, {van der Tak}, {van
  der Wiel}, {Viti}, {Wyrowski}, \& {Yorke}}]{HilyBlant10}
{Hily-Blant}, P., {Maret}, S., {Bacmann}, A., {et~al.} 2010, \aap, 521, L52+

\bibitem[{Hily-Blant {et~al.}(2005)Hily-Blant, Pety, \&
  Guilloteau}]{HilyBlant05}
Hily-Blant, P., Pety, J., \& Guilloteau, S. 2005, CLASS evolution: I. Improved
  OFT support, Tech. rep., IRAM

\bibitem[{{Johansson} {et~al.}(1984){Johansson}, {Andersson}, {Ellder},
  {Friberg}, {Hjalmarson}, {Hoglund}, {Irvine}, {Olofsson}, \&
  {Rydbeck}}]{Johansson84}
{Johansson}, L.~E.~B., {Andersson}, C., {Ellder}, J., {et~al.} 1984, \aap, 130,
  227

\bibitem[{{M{\" u}ller} {et~al.}(2001){M{\" u}ller}, {Thorwirth}, {Roth}, \&
  {Winnewisser}}]{Muller01}
{M{\" u}ller}, H.~S.~P., {Thorwirth}, S., {Roth}, D.~A., \& {Winnewisser}, G.
  2001, \aap, 370, L49

\bibitem[{{Moreau} {et~al.}(2008){Moreau}, {Dubernet}, \&
  {M{\"u}ller}}]{Moreau08}
{Moreau}, N., {Dubernet}, M.~L., \& {M{\"u}ller}, H. 2008, in Astronomical
  Spectroscopy and Virtual Observatory, ed. {M.~Guainazzi \& P.~Osuna}, 195--+

\bibitem[{Pickett {et~al.}(1998)Pickett, Poynter, Cohen, Delitsky, Pearson, \&
  M\"uller}]{Pickett98}
Pickett, H.~M., Poynter, R.~L., Cohen, E.~A., {et~al.} 1998, JQSRT, 60, 830

\bibitem[{{Pilbratt} {et~al.}(2010){Pilbratt}, {Riedinger}, {Passvogel},
  {Crone}, {Doyle}, {Gageur}, {Heras}, {Jewell}, {Metcalfe}, {Ott}, \&
  {Schmidt}}]{Pilbratt10}
{Pilbratt}, G.~L., {Riedinger}, J.~R., {Passvogel}, T., {et~al.} 2010, \aap,
  518, L1+

\bibitem[{Salgado {et~al.}(2009)Salgado, Osuna, Osuna, Barbarisi, Dubernet, \&
  Tody}]{Salgado09}
Salgado, J., Osuna, P., Osuna, M., {et~al.} 2009, Simple Line Access Protocol,
  Tech. rep., International Virtual Observatory Alliance

\bibitem[{{Schilke} {et~al.}(2001){Schilke}, {Benford}, {Hunter}, {Lis}, \&
  {Phillips}}]{Schilke01}
{Schilke}, P., {Benford}, D.~J., {Hunter}, T.~R., {Lis}, D.~C., \& {Phillips},
  T.~G. 2001, \apjs, 132, 281

\bibitem[{{Schilke} {et~al.}(1997){Schilke}, {Groesbeck}, {Blake}, \&
  {Phillips}}]{Schilke97a}
{Schilke}, P., {Groesbeck}, T.~D., {Blake}, G.~A., \& {Phillips}, T.~G. 1997,
  \apjs, 108, 301

\bibitem[{{van Dishoeck} {et~al.}(1995){van Dishoeck}, {Blake}, {Jansen}, \&
  {Groesbeck}}]{vanDishoeck95}
{van Dishoeck}, E.~F., {Blake}, G.~A., {Jansen}, D.~J., \& {Groesbeck}, T.~D.
  1995, \apj, 447, 760+

\bibitem[{Wang {et~al.}(2010)Wang, Bergin, Crockett, Bell, Blake, Caux, \&
  co.}]{Wang10}
Wang, S., Bergin, E., Crockett, N., {et~al.} 2010, \aap, in press

\bibitem[{{Wootten}(2008)}]{Wooten08}
{Wootten}, A. 2008, \apss, 313, 9

\end{thebibliography}

\end{document}